\documentclass[preprint,showpacs,preprintnumbers,amsmath,amssymb]{revtex4}

\usepackage{graphicx}
\usepackage{dcolumn}
\usepackage{bm}
\usepackage{color}
\usepackage{ulem}

\newcommand{\bra}[1]{\ensuremath{\left\langle#1\right|}}
\newcommand{\ket}[1]{\ensuremath{\left|#1\right\rangle}}

\begin{document}

\title{Measurement of paramagnetic spin concentration in a solid-state system using double electron-electron resonance} 

\author{Viktor Stepanov}
\affiliation{Department of Chemistry, University of Southern California, Los Angeles CA 90089, USA}

\author{Susumu Takahashi}
\email{susumu.takahashi@usc.edu}
\affiliation{Department of Chemistry, University of Southern California, Los Angeles CA 90089, USA}
\affiliation{Department of Physics and Astronomy, University of Southern California, Los Angeles CA 90089, USA}


\begin{abstract}
Diamond has been extensively investigated recently due to a wide range of potential applications of nitrogen-vacancy (NV) defect centers existing in a diamond lattice.
The applications include magnetometry and quantum information technologies, and long decoherence time ($T_2$) of NV centers is critical for those applications.
Although it has been known that $T_2$ highly depends on the concentration of paramagnetic impurities in diamond,
precise measurement of the impurity concentration remains challenging.
In the preset work, we show a method to determine a wide range of the nitrogen concentration ($n$) in diamond using a wide-band high-frequency electron spin resonance and double electron-electron resonance spectrometer.
Moreover, we investigate $T_2$ of the nitrogen impurities and show the relationship between $T_2$ and $n$.
The method developed here is applicable for various spin systems in solid and implementable in nanoscale magnetic resonance spectroscopy with NV centers to characterize the concentration of the paramagnetic spins within a microscopic volume.
\end{abstract}


\maketitle

\section{Introduction}
A nitrogen-vacancy center (NV) in diamond is a promising candidate for investigation of spin physics~\cite{Gruber97, Jelezko02} and applications to quantum information processing~\cite{JelezkoGate04, Jiang09, Neumann10} and quantum nanoscale sensing~\cite{Degen08, Balasubramanian08, Maze08, Taylor08, Steinert13, Kaufmann13, Mamin13sci, Staudacher13, Ohashi13, muller14, Maletinsky12, Abeywardana14} because of its remarkable properties including excellent photostability and capability to detect a single NV center at room temperature~\cite{Gruber97}.
For the fundamental sciences and applications, long coherence of a NV center is critical.
Coherence of a NV center highly depends on contents of paramagnetic impurities in diamond.
In particular, nitrogen related impurities including well-known single substitutional nitrogen impurities (N spins, also known as P1 centers) are often abundant in many diamond crystals.
For example, type-Ib and type-IIa diamonds typically contain nitrogen impurity concentration in the range of 10$-$100 parts-per-million (ppm) and tens of parts-per-billion (ppb), respectively.
Coherence in such diamond crystals are largely affected by the concentration of nitrogen impurities~\cite{Vanwyk97, Wang13}.

Moreover, for past several years, ensembles of NV centers of high concentrations($\sim$1$-$100 ppm) have taken a rapidly growing interest to study and fabricate ~\cite{Jarmola12, Wolf15, Grezes15}, showing that precise determination of the concentration of NV centers and N spins in diamond is highly useful.
Unfortunately, currently available techniques have several limitations.
For example, infrared absorption spectroscopy is a commonly-used technique to determine N spin concentration, however the sensitivity is often not high enough to measure type-IIa diamond~\cite{Iakoubovskii00}.
Lineshape analysis of electron spin resonance (ESR) spectroscopy has also been used to estimate the N spin concentrations,
however accuracy of the analysis relies on a well-calibrated reference sample~\cite{Vanwyk97}.

In this article, we propose and demonstrate a method to determine the concentration of paramagnetic impurities in solid-state systems with high precision and no reference sample using 115 GHz double electron-electron resonance (DEER) spectroscopy at room temperature.
DEER spectroscopy is known to be a powerful technique to probe the magnetic dipole interaction between paramagnetic spins.
For the investigation, we employ a home-built high-frequency (HF) ESR/DEER spectrometer with capability to output in the frequency range of 107$-$120 GHz so that the system enables to perform high spectral resolution ESR/DEER spectroscopy with different groups of spins.
First, we measure ESR spectrum of paramagnetic spins in diamond which allows us to identify a type of impurities.
The ESR spectrum analysis confirms that a majority of paramagnetic spins in both type-Ib and type-IIa diamonds are N spins.
Then we perform pulsed ESR experiment to determine spin decoherence time ($T_2$) in the diamond crystals.
Moreover we perform DEER spectroscopy to determine the concentration of N spins in the range of 0.1$-$100 ppm.
Finally, we investigate the relationship between the concentration of N spins and their spin decoherence time ($T_2$).

\section{Experiment}
For the investigation, we employed several synthetic diamond crystals including type-Ib and type-IIa crystals from DiAmante Industries, LLC~\cite{DiAmante}, Element 6~\cite{E6} and Sumitomo Electric~\cite{Sumitomo}.
The investigation was performed using a home-built 115 GHz ESR/DEER spectrometer. The 115 GHz ESR system employs a high-power ($\sim$700 mW) solid-state source, quasioptical bridge, a corrugated waveguide and a 12.1 T cryogenic-free superconducting magnet.
The detection system is based on the induction mode detection to measure in-phase and quadrature components of ESR signals.
The system also has a wide-band DEER capability ($\sim$13 GHz) which is required for the present study.
Details of the system have been described elsewhere~\cite{Cho14, Cho15}.

\subsection{Spin echo measurement}
\begin{figure}
\includegraphics[width=70 mm]{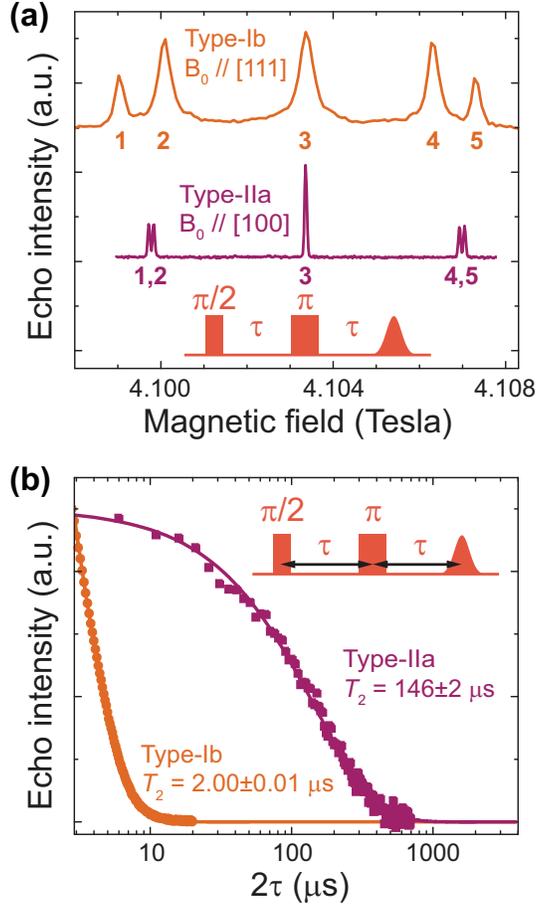}
\caption{SE measurements of type-Ib and type-IIa diamond crystals.
(a) SE intensity as a function of magnetic fields. The applied pulse sequence is shown in the inset. In the measurement of the type-Ib diamond, The durations of $\pi$/2 and $\pi$ pulses were 150 ns and 250 ns and $\tau$ was 1.5 $\mu$s. The data was taken with 32 averages with 20 ms of the repetition time. In the measurement of the type-IIa diamond, the durations of the $\pi$/2 and $\pi$ pulses were 250 ns and 450 ns and $\tau$ was 3 $\mu$s. The data was taken with 256 averages with 20 ms of the repetition time. The magnetic field was applied along the $[111]$ direction for type-Ib crystals and the $[100]$ direction for type-IIa.
(b) SE intensity as a function of $\tau$ to measure spin decoherence time $T_2$. The decays of the SE were fitted by a single exponential function to extract $T_2$ (solid lines). The data of the type-Ib (type-IIa) diamond was taken with 128 (256) averages.}
\end{figure}
Figure 1 shows 115 GHz ESR measurements of type-Ib and type-IIa diamond crystals performed by monitoring the spin echo (SE) intensity as a function of magnetic fields.
The type-Ib diamond crystal has a polished face normal to the $[111]$ crystallographic axis while the type-IIa diamond crystal has a polished face normal to the $[100]$ axis.
In both measurements, the magnetic field was applied perpendicular to the polished surface.
As shown in Fig.~1a, the ESR spectrum of the type-Ib diamond sample shows five pronounced peaks representing N spins ($\hat{H}_N = g \mu_B \vec{B} \hat{S} + \hat{S} \overset{\text{\tiny$\leftrightarrow$}}{A} \hat{I}$, $S$ = 1/2, $g=2.0024$, $I$ = 1, $A_{x,y}$ = 82 MHz, and $A_z$ = 114 MHz).
These five peaks originate (labeled as 1, 2, 3, 4, and 5) from the four principle axes of N spins, {\it i.e.}, $[111]$, $[11\bar{1}]$, $[1\bar{1}1]$ and $[\bar{1}11]$, and the hyperfine interaction to $^{14}$N nuclear spin~\cite{Loubser78, Takahashi08}.
The intensity of the ESR signals represents the population of each group, with the population ratio corresponding to $1:3:4:3:1$ for Group 1$-$5, respectively.
In addition, we measured the SE intensity of the N spins as a function of magnetic fields in the type-IIa diamond.
As shown in Fig. 1a, the width of the observed signals were significantly narrower than those of the type-Ib crystal.
Next, figure~1b shows spin decoherence time ($T_2$) measurements of the type-Ib and type-IIa samples.
We observed that the SE decayed exponentially as a function of 2$\tau$ in both cases.
As indicated in Fig.~1b, $T_2$ for the type-IIa diamond was nearly two orders of the magnitude longer than that of the type-Ib diamond
while both samples have similar spin-lattice relaxation times ($T_1$) of several ms (data not shown).
We also found that $T_2$ values of all groups were very similar.

\subsection{Double electron-electron resonance spectroscopy}
\begin{figure}
\includegraphics[width=70 mm]{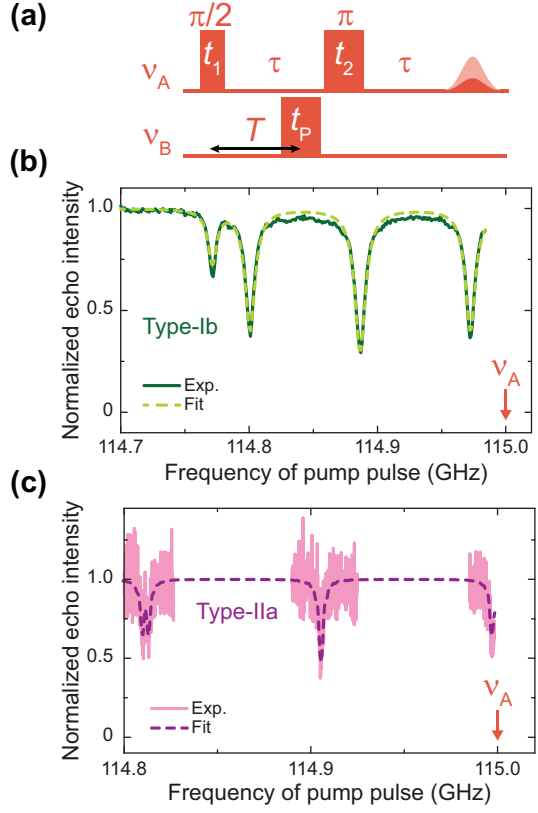}
\caption{DEER spectroscopy of the type-Ib and type-IIa diamond crystals. (a) three-pulse DEER sequence used in the experiment, where $t_1$ and $t_2$ denote duration of $\pi/2$ and $\pi$ pulses for A spins, respectively, $t_p$ and $T$ duration and delay of $\pi$ pulse for B spins.
(b)$\&$(c) DEER spectrum of N spins in type-Ib and type-IIa diamonds, respectively.
The DEER signals were normalized by the SE signals.
Experimental parameters were $t_1$ = 250 ns, $t_1$ = 450 ns, $t_p$ = 450 ns, $\tau$ = 2.5 $\mu$s, $T$ = 2 $\mu$s in case of type-Ib diamond, and $t_1$ = 250 ns, $t_1$ = 450 ns, $t_p$ = 450 ns, $\tau$ = 110 $\mu$s, $T$ = 109.45 $\mu$s in case of type-IIa diamond.
The data of the type-Ib (type-IIa) diamond was taken with 128 (256) averages.
Purple and brown dashed lines represent the best fit of experimental data using Eqn.~\ref{eq:deer}.
}
\end{figure}
Next, we performed DEER spectroscopy to probe the magnetic dipole interaction between N spins.
For DEER spectroscopy of the type-Ib diamond, the N spins at $B_0$ = 4.099 Tesla (Group 1), whose axis is along $[111]$ and whose nuclear spin state is $|m_I = 1\rangle$, were used as probe spins (A spins).
B spins (other N spins in Group 2$-$5 in Fig. 1a) were used as pump spins.
Then we applied the three-pulse DEER sequence to probe the magnetic dipolar coupling between N spins in diamond~\cite{Milov81}.
As shown in the inset of Fig. 2a, the applied DEER sequence consisted of the SE sequence for A spins at the frequency of $\nu_A$ = 115 GHz and a single $\pi$ pulse for B spins at the frequency of $\nu_B$.
In the DEER spectroscopy, changes in the SE signal occur when the effective magnetic dipolar fields at A spins are altered by B spins that are flipped by the $\pi$ pulse.
As shown in Fig.~2b, four DEER signals of N spins were clearly observed as reductions of the SE intensity of A spins.
The signals were centered at 114.772, 114.801, 114.886 and 114.971 GHz, corresponding to B spins in Group 2, 3, 4, and 5, respectively.
Thus, the result confirms direct observation of the dipolar coupling between N spins in the type-Ib diamond.
Similarly, we performed the DEER measurement with the type-IIa diamond, and,
as shown in Fig.~2c, observed the DEER signals.

\section{model}
\subsection{Spin echo}
There exist several processes which can contribute to the SE decay, including the spin flip-flops of N spin bath, the instantaneous diffusion, $^{13}$C nuclear spins and the single spin flips ($T_1$ process).
As reported previously, the spin flip-flop (also known as the spectral diffusion) is one of the major decoherence sources in type-Ib diamond crystals~\cite{Takahashi08, Vanwyk97}.
The spin flip-flop process causes dipolar-field fluctuations at the sites of the excited spins and
the decoherence rate of this process linearly depends on the concentration of surrounding non-excited N spin bath~\cite{Wang13, Vanwyk97}.
On the other hand, in the case of type-IIa, it has been shown that the nuclear spin decoherence is pronounced~\cite{Childress06, Gaebel06}.
In addition, the SE decay may be speeded up by the process of instantaneous diffusion that manifests itself upon application of $\pi$ pulse due to dipole-dipole interactions between the excited spins.
In the case of the instantaneous diffusion process, the SE decay depends on the concentration of the excited spins,
therefore the contribution of the instantaneous diffusion will be varied between spin groups with different concentrations of N spins, {\it e.g.} group 1 and 3 in Fig.~1a.
However our observation of similar $T_2$ times between different groups indicates that the instantaneous diffusion is insignificant in our experiments.
Moreover, because the observed $T_1$ is much longer than $T_2$, the $T_1$ process is negligible in the present case.

Next, we discuss the SE decay to estimate the spin flip-flop rate with the use of a model for the dipolar-coupled spins developed in Ref.~\cite{Salikhov80}.
According to Ref.~\cite{Salikhov80}, the SE decay due to the spectral diffusion is described by the following expression,
\begin{equation}
\label{eq:a}
SE(2\tau)=\exp\left(
- n \int\displaylimits_{0}^{\infty} f(W,W_{max}) \int\displaylimits_{V} \big[1 - v_0(2\tau,W)\big] \, \text{d}V \, \text{d}W
\right),
\end{equation}
where $W$ is the rate of the spin flip-flops of bath spins.
$v_0$ represents SE signals of a single excited spin dipolar-coupled to a non-excited bath spin with the relative radius vector ($\vec{r}(r,\theta)$),
which is given by,
\begin{equation*}
v_0(2\tau, W)= \left[
\left(\text{cosh} R\tau + \frac {W}{R} \text{sinh}R \tau\right)^2 + \frac {A^2}{4 R^2} \text{sinh}^2 R \tau
\right] \exp{\left(- 2 W \tau\right)},
\end{equation*}
where $A\equiv \mu_0 \mu_B^2 g_1 g_2(1 - 3 \cos^2 \theta)/(4 \pi \hbar r^3)$ and $R^2 \equiv W^2 - \frac{1}{4}A^2$.
$\mu_0$ is the vacuum permeability, $\mu_B$ is the Bohr magneton, $\hbar$ is the reduced Planck constant, $g_1$ and $g_2$ are $g$-factors of the excited and bath spins, respectively.
The integration over the sample volume $V$ in Eqn.~(\ref{eq:a}) takes into account all possible $r$ and $\theta$.
The integration over $W$ accounts for a distribution of the flip-flop rate within the sample where the distribution function $f(W,W_{max})$ is given by~\cite{Salikhov80},
\begin{equation}
\label{eq:b}
f(W,W_{max}) = \sqrt{\frac{3 W_{max}}{2 \pi W^3}} \exp{\left(- \frac{3 W_{max}}{2 W}\right)}.
\end{equation}
where $f(W,W_{max})$ is maximum at the flip-flop rate of $W=W_{max}$.
\begin{figure}
\includegraphics[width=150 mm]{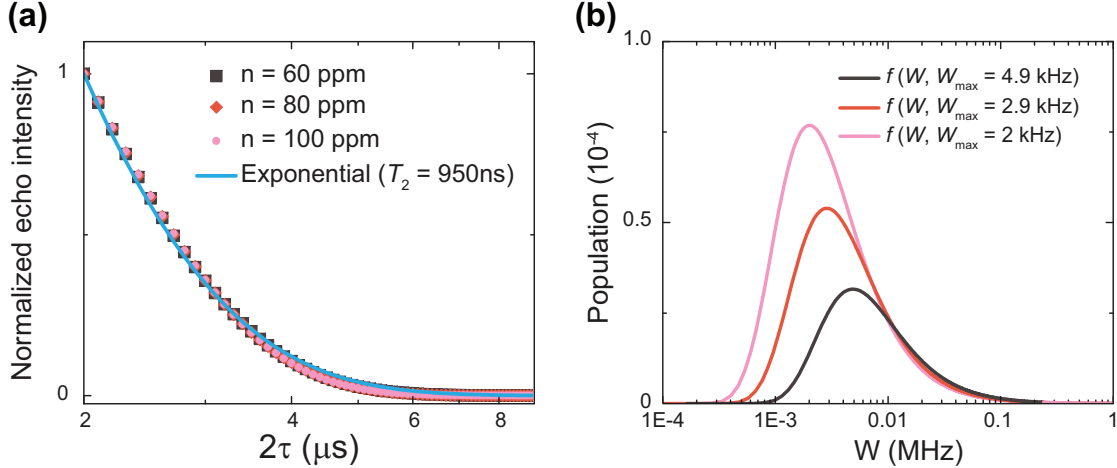}
\caption{(a) Analyses of a single exponential SE decay with $T_2$ = 950 ns (cyan) using  Eqn.~(\ref{eq:a}). 4.9, 2.9 and 2 kHz of $W_{max}$ were obtained from the fits for 60 (red), 80 (green) and 100 (blue) ppm of N concentrations, respectively.
(b) Flip-flop rate distribution among N spins obtained using  Eqn.~(\ref{eq:b}) for 4.9 (red), 2.9 (green) and 2 kHz (blue) of $W_{max}$.}
\label{fig:a}
\end{figure}
Using the model above, we estimate an average flip-flop rate of N spins in diamond.
We first considered a single exponential SE decay with $T_2$ = 950 ns ($\sim$the shortest $T_2$ observed in our experiments)
and performed a fit using Eqn.~(\ref{eq:a}) with a fixed N concentration to extract $W_{max}$.
As shown in Fig.~3a, the SE model (Eqn.~(\ref{eq:a})) fits well with a single exponential decay with $T_2$ = 950 ns and
the fit results give $\sim$4.9, $\sim$2.9, and $\sim$2 kHz of $W_{max}$ for 60, 80 and 100 ppm of the concentrations, respectively.
The flip-flop distribution function (Eqn.~\ref{eq:b}) for the obtained $W_{max}$ are plotted in Fig.~3b.
As shown in Fig.~\ref{fig:a}b, a major population of the flip-flop rate ranges from $\sim$1 kHz to $\sim$1 MHz.
In addition, an average flip-flop rate is given by,
\begin{equation*}
{\langle W \rangle}_{80 \%} = \left. \left[ \sqrt{\frac{6 W_{max} b}{\pi}} \exp{\left(-\frac{3 W_{max}}{2b}\right)} - 3 W_{max} \text{erfc}\left( \sqrt{\frac{3 W_{max}}{2b}} \right) \right] \right|_{b=50 W_{max}}
\approx 7.1 W_{max}
\end{equation*}
where the upper limit of the integration was set at 50 $W_{max}$ (corresponding to 80$\%$ of the cumulative percentage) to avoid the divergence of the integral to evaluate the $\langle W \rangle$. Using values for $\langle W \rangle_{80\%}$, the average flip-flop events $2 \tau \langle W \rangle_{80\%}$ during the DEER sequence (2$\tau$ = 3 $\mu$s for the sample with the shortest $T_2$) were estimated as 0.1, 0.06 and 0.04 for 60, 80 and 100 ppm, respectively.
Moreover, for longer $T_2$ times, the flip-flop probability is expected to be even lower.
Given the small flip-flop probability on the time scale of the DEER experiment, we consider the N spins to be in the static regime to model the DEER signal.

\subsection{Double electron-electron resonance}
\begin{figure}
\includegraphics[width=100 mm]{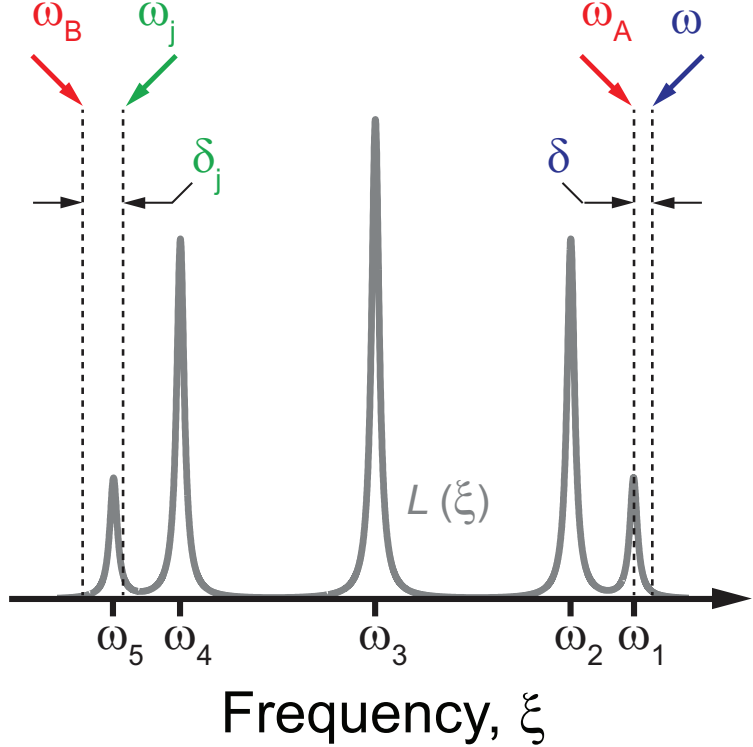}
\caption{Schematics for the DEER model. $L(\xi)$ is the lineshape function. ${\omega}_m$ is the center frequency of Group $m$ ($m$=1$-$5).
$\omega_A$ and $\omega_B$ are microwave frequencies of the probe and pump pulses, respectively. $\omega$ and $\omega_j$ are the Larmor frequencies of A and B spins, respectively. $\delta$ and $\delta_j$ are frequency offsets of A and B spins from the pump and probe frequencies, respectively. A and B spins were chosen close to the probe and pump frequencies, to indicate, that spins can be excited by a respective pulse with a small frequency offsets. However, in general, as in our consideration, they can be anywhere within the lineshape $L$.}
\end{figure}
In this section, we model DEER signals for ensemble N spins.
The DEER signal is produced by probe N spins (A spins) interacting with resonant N spins to the pump pulse (B spins) and the rest of spins in diamond (C spins).
C spins include both non-resonant N spins and nuclear spins.
The center frequencies of ESR transitions of N spins are given by the Hamiltonian of N spins ($\hat{H}_N = g \mu_B \vec{B} \hat{S} + \hat{S} \overset{\text{\tiny$\leftrightarrow$}}{A} \hat{I}$).
Moreover, all ESR transitions have equal linewidths due to randomly distributed N and nuclear spins in the diamond lattice, giving rise to inhomogeneously broadened spectral lines ({\it e.g.} Group 1$-$5 in Fig.~4).
We describe each spectral line by Lorentzian lineshape with a half width of $\Delta \omega$.
Thus the total lineshape is given by $L(\xi) = \frac{1}{\pi} \sum_{m} f_m \frac{\Delta \omega}{\Delta \omega^2 + (\xi - \omega_m)^2}$, where $f_m$ and $\omega_m$ being fraction of spins and transition frequency of Group $m$, respectively.

Here, we focus on the case when magnetic field is applied along the [111] direction and a DEER lineshape is shown in Fig.~4.
We start by considering a single A spin with the Larmor frequency $\omega$ (see Fig.~4) as a two-level system (TLS) represented by Hamiltonian in units of frequency, $\hat{H}_0 = \omega \hat{S}_z$.
During application of the probe pulse with microwave frequency $\omega_A$, applied at the center frequency of  Group 1 ($\omega_1$), the total Hamiltonian is given by $\hat{H}=\hat{H}_0 + \hat{H}_{MW} = \omega \hat{S}_z + 2 \Omega \hat{S}_x \cos{\omega_A t}$, where $\Omega = g \mu_B b_1 / \hbar$ and $b_1$ is the strength of the microwave field.
The frequency offset ($\delta$) in Fig.~4, defined as $\delta \equiv \omega - \omega_A$, is due to local magnetic fields from B and C spins, $\it{ i.e.}$ $\delta = g \mu_B (b_B + b_C)$, where $b_B(t) = \sum_{j} b_j (t)$ and $b_C(t) = \sum_{k} c_k (t)$, $j$ and $k$ are indexes of B and C spins, and $b_j(t)$ and $c_k(t)$ are magnetic fields produced by $j$-th B and $k$-th C spins at a single A spin, respectively.
Due to the low probability of the flip-flop as discussed in Sec.~III-A, $\delta$ is considered to be time-independent.
Moreover, to calculate DEER signals below, we assume $|g \mu_b b_b/\hbar| \ll |\delta|$ for A spins contributing to SE signals in DEER experiment because of the low concentration ($<$ $\sim$10$^{19}$ spins/cm$^3$) and partial excitation of N spins.
$|g \mu_b b_b/\hbar| \ll |\delta|$ is also commonly employed in dilute spin systems ($<$ 10$^{20}$ spins/cm$^3$)~\cite{Salikhov76}.
The above assumptions ensure constant $\delta$ during DEER sequence.

First, we calculate SE signal produced by a single A spin during the pulse sequence ($t_1-\tau-t_2-\tau$).
The spin state  by the end of the sequence ($\ket{\psi_{2 \tau}}$) is given by,
\begin{equation}
\label{eq:c}
\ket{\psi_{2 \tau}} = \hat{U}_2(\tau)\hat{R}(t_2) \hat{U}_1(\tau) \hat{R}(t_1) \ket{\psi_0},
\end{equation}
where $\ket{\psi_0}$ is the initial state. $\hat{R}(t_i) \equiv \exp{\left[ - i \left(\delta \hat{S}_z + \Omega \hat{S}_x \right) t_i \right]}$ is a propagator that describes evolution of TLS under the microwave excitation in the rotating frame with the microwave frequency ($\omega_A$).
In a matrix representation in the basis of  \ket{+} and \ket{-} states, $\hat{R}(t_i)$ is given by,
\begin{equation*}
\hat{R}(t_i) =
\left(\begin{matrix}
c_i - i \frac{\delta}{\Omega_A} s_i & -i \frac{\Omega}{\Omega_A} s_i \\
-i \frac{\Omega}{\Omega_A} s_i & c_i + i \frac{\delta}{\Omega_A} s_i
\end{matrix} \right),
\end{equation*}
where $\Omega_A \equiv \sqrt{\delta^2 + \Omega^2}$, $c_i \equiv \cos{\Omega_A t_i/2}$ and $s_i \equiv \sin{\Omega_A t_i/2}$.
$U_i$ is a free evolution propagator defined as
\begin{equation*}
\hat{U}_i(\tau) =
\left( \begin{matrix}
e^{-i (\varphi_i + \phi_i)/2} & 0 \\
0 & e^{i (\varphi_i + \phi_i)/2}
\end{matrix} \right),
\end{equation*}
with $\varphi_1 \equiv \frac{g \mu_B}{\hbar} \int\displaylimits_{0}^{\tau} b_B(t) \, \text{d}t$, $\varphi_2 \equiv \frac{g \mu_B}{\hbar} \int\displaylimits_{\tau}^{2\tau} b_B(t) \, \text{d}t$, $\phi_1 \equiv \frac{g \mu_B}{\hbar} \int\displaylimits_{0}^{\tau} b_C(t) \, \text{d}t$ and $\phi_2 \equiv \frac{g \mu_B}{\hbar} \int\displaylimits_{\tau}^{2\tau} b_C(t) \, \text{d}t$

Using Eqn.~(\ref{eq:c}), the magnetic field component in the rotating frame along $y$-axis of a single A spin with the initial state $\ket{\psi_0} = \ket{-}$, is calculated as
\begin{equation*}
\begin{split}
\big \langle \hat{S}_y \big \rangle_{s} & = \bra{\psi_{2 \tau}} \hat{S}_y \ket{\psi_{2 \tau}} \\
& = \left[ \frac{\Omega}{\Omega_A} c_1 s_1 c_2^2 - \frac{\delta^2 \Omega}{\Omega_A^3} (c_1 s_1 s_2^2 + 2 s_1^2 c_2 s_2) \right] \cos 2 \delta \tau \\
& + \left[ \frac{\delta^3 \Omega}{\Omega_A^4} s_1^2 c_2^2 - \frac{\delta \Omega}{\Omega_A^2} (2 c_1 s_1 c_2 s_2 + s_1^2 c_2^2) \right] \sin 2 \delta \tau \\
& + \frac{\Omega}{\Omega_A^3} c_2 s_2 \left[ \delta^2 +\Omega^2 (c_1^2 - s_1^2) \right] \cos \delta \tau \\
& - \frac{\delta \Omega}{\Omega_A^4} s_2^2 \left[ \delta^2 +\Omega^2 (c_1^2 - s_1^2) \right] \sin \delta \tau \\
& + \left[ - \frac{\Omega^3}{\Omega_A^3} c_1 s_1 s_2^2 \right] \cos \left( \varphi_1 - \varphi_2 \right)
+ \left[ \frac{\delta \Omega^3}{\Omega_A^4} s_1^2 s_2^2 \right] \sin \left( \varphi_1 - \varphi_2 \right).\\
\end{split}
\end{equation*}
After omitting the FID signals that are averaged out on the time scale of $T_2^*$~\cite{Salikhov80},
the $\big \langle \hat{S}_y \big \rangle_{s}$ is reduced to
\begin{equation}
\label{eq:sy}
\big \langle \hat{S}_y \big \rangle_{s} \approx
\left[ - \frac{\Omega^3}{\Omega_A^3} c_1 s_1 s_2^2 \right] \cos \left(\varphi_1 - \varphi_2 \right)
+ \left[ \frac{\delta \Omega^3}{\Omega_A^4} s_1^2 s_2^2 \right] \sin \left( \varphi_1 - \varphi_2 \right).
\end{equation}
Similarly, $\big \langle \hat{S}_x \big \rangle_{s}$ in the rotating frame is found as
\begin{equation}
\label{eq:e}
\big \langle \hat{S}_x \big \rangle_{s} \approx
\left[ - \frac{\Omega^3}{\Omega_A^3} c_1 s_1 s_2^2 \right] \sin \left( \varphi_1 - \varphi_2 \right)
+ \left[ - \frac{\delta \Omega^3}{\Omega_A^4} s_1^2 s_2^2 \right] \cos \left( \varphi_1 - \varphi_2 \right).
\end{equation}

Next, the SE signals of a single A spin in the DEER measurement is calculated.
When the pump pulse with the frequency ($\omega_B$) excites B spins,
the phase accumulated by the A spin during 2$\tau$ is expressed as
\begin{equation}
\label{eq:f}
\delta \varphi \equiv \varphi_1 - \varphi_2 = \frac{g \mu_B}{\hbar} \sum_{j} \left( b_j (T - t_p/2) + \int\displaylimits_{0}^{t_p} b_j^{MW}(t) \, \text{d}t + b_j^{MW}(t_p) \left[ (\tau - T - t_p/2) - \tau \right] \right),
\end{equation}
where $b_j \equiv \mu_0 \mu_B g_B (3 \cos^2 \theta_j - 1) \sigma_j /(4 \pi \hbar r_j^3)$ is a magnetic field produced by the $j$-th B spin at the A spin before the pump pulse is applied.
$\sigma_j$ is the spin state of the $j$-th B spin ($\sigma_j \pm 1/2$).
$\vec{r}_j (r_j, \theta_j)$ is the radius vector of the dipole interaction between the $j$-th B spin and the A spin.
$b_j^{MW} = b_j \left[ \delta_j^2 + \Omega^2 (c_j^2 - s_j^2)\right]/\Omega_{B,j}^2$ with $\delta_j \equiv \omega_B -\omega_j$ ($\omega_j$ is the Larmor frequency of the $j$-th B spin. See Fig.~4), $\Omega_{B,j} \equiv \sqrt{\delta_j^2 + \Omega^2}$, $c_j \equiv \cos \Omega_{B,j}t/2$ and $s_j \equiv \sin \Omega_{B,j}t/2$.
It is important to note that Eqn.~(\ref{eq:f}) takes into account off-resonant excitation of the B spins which is represented by ($\sigma_j$, $r_j$, $\theta_j$) and $\delta_j$.
Moreover, Eqn.~(\ref{eq:f}) can be further simplified in the present case ($t_p \ll 2\tau$ and $T \sim \tau$) to give
\begin{equation*}
\delta \varphi \approx \frac{\mu_0}{4 \pi} \frac{\mu_B^2 g_A g_B (2 T)}{\hbar}
\sum_{j} \frac{\Omega^2}{\delta_j^2 + \Omega^2} \sin^2 \left( \sqrt{\delta_j^2 + \Omega^2} \frac{t_p}{2} \right)
\frac{(3 \cos^2 \theta_j - 1) \sigma_j }{r_j^3}.
\end{equation*}
Using the approach described in Ref.~\cite{Cho15, Mims68, Salikhov76}, the SE signal ($\big \langle \hat{S}_y \big \rangle_{s}$ and $\big \langle \hat{S}_x \big \rangle_{s}$) is averaged over B spins ($r_j$, $\theta_j$, $\sigma_j$, $\delta_j$),
\begin{equation}
\label{eq:g}
\Big \langle \big \langle \hat{S}_y \big \rangle_{s} \Big \rangle_B \approx \left[ - \frac{\Omega^3}{\Omega_A^3} c_1 s_1 s_2^2 \right]
\exp \left(- \frac{2 \pi \mu_0 \mu_B^2 g_A g_B T}{9\sqrt{3}\hbar} n \Big \langle \sin^2 \frac{\theta}{2} \Big \rangle_L \right),
\end{equation}
and
\begin{equation}
\label{eq:h}
\Big \langle \big \langle \hat{S}_x \big \rangle_{s} \Big \rangle_B \approx \left[ - \frac{\delta \Omega^3}{\Omega_A^4} s_1^2 s_2^2 \right]
\exp \left(- \frac{2 \pi \mu_0 \mu_B^2 g_A g_B T}{9\sqrt{3}\hbar} n \Big \langle \sin^2 \frac{\theta}{2} \Big \rangle_L \right),
\end{equation}
where $\langle \sin^2 \frac{\theta}{2} \rangle_L \equiv \int\displaylimits_{- \infty}^{+ \infty}
\frac{\Omega^2}{(\xi - \omega_B)^2 + \Omega^2} \sin^2 \left( \sqrt{(\xi - \omega_B)^2 + \Omega^2} \frac{t_p}{2} \right) L(\xi)\, \text{d}\xi$.

To calculate DEER signal components in the rotating frame ($I_x$ and $I_y$) produced by an ensemble of A spins, the DEER signals are first obtained for a single A spin with the $\ket{\psi} = \ket{+}$ initial spin state, similarly to above calculations, and averaged over $\ket{+}$ and $\ket{-}$ spin states with the use of thermal populations in each state, resulting in the thermal magnetization factor ($\Delta \equiv \text{tanh} (\hbar \omega_A/2 k_B T_0)$ where $T_0$ is sample temperature) for Eqns.~(7) and (8).
Next, the signals are averaged over the lineshape ($L$) to give
\begin{equation*}
I_y = \Delta~\Big \langle-\frac{\Omega^3}{\Omega_A^3} c_1 s_1 s_2^2 \Big \rangle_L
\exp \left(- \frac{2 \pi \mu_0 \mu_B^2 g_A g_B T}{9\sqrt{3}\hbar} n \Big \langle \sin^2 \frac{\theta}{2} \Big \rangle_L \right),
\end{equation*}
and
\begin{equation*}
I_x = \Delta~\Big \langle - \frac{\delta \Omega^3}{\Omega_A^4} s_1^2 s_2^2 \Big \rangle_L
\exp \left(- \frac{2 \pi \mu_0 \mu_B^2 g_A g_B T}{9\sqrt{3}\hbar} n \Big \langle \sin^2 \frac{\theta}{2} \Big \rangle_L \right).
\end{equation*}
where $\left\langle...\right\rangle_{L}$ represents averaging over the inhomogeneous lineshape $L$.
The latter being averaged out to zero when the probe frequency is centered with Group 1, thus the DEER intensity ($I_{\Omega}$) is given by
\begin{equation}
\label{eq:i}
\begin{split}
I_{\Omega} & \equiv \sqrt{I_x^2 + I_y^2} = I_y \\
& = \Delta~\Big \langle \frac{\Omega^3}{\Omega_A^3} c_1 s_1 s_2^2 \Big \rangle_L
\exp \left(- \frac{2 \pi \mu_0 \mu_B^2 g_A g_B T}{9\sqrt{3}\hbar} n \Big \langle \sin^2 \frac{\theta}{2} \Big \rangle_L \right) \exp \left( - \frac{2 \tau}{T_2}\right),
\end{split}
\end{equation}
where the SE decay ($ \exp \left( - 2 \tau/T_2\right)$) was added.
In the case where the excitation bandwidth is larger that the inhomogeneous line ($\delta \ll \Omega$, then $\langle \sin^2 \frac{\theta}{2} \rangle_L$ = 1), Eqn.~(\ref{eq:i}) reduces to the result obtained previously~\cite{Milov81, Cho15}:
\begin{equation*}
I_{DEER}(n) \sim \exp\left(-\frac{2 \pi \mu_0 \mu_B^2 g_A g_B T}{9\sqrt{3}\hbar} n\right).
\end{equation*}
Furthermore, the obtained $\langle \sin^2 \frac{\theta}{2} \rangle_L$ function in Eqn.~(\ref{eq:i}) has been previously considered in the context of instantaneous diffusion~\cite{Salikhov80, Gelardi01} and DEER background signals in stabilized radical systems~\cite{Milov84}.
In addition, the SE intensity was calculated previously without fully taking into account the off-resonant excitation~\cite{Salikhov80, Salikhov76}.
In general, the off-resonant excitation not only reduces the tipping angle, but also results in the finite spin projection along the microwave field
that was not considered in the previous models, however, in the present case, this contribution is critical.

In the present experiment, the microwave power is distributed across the sample,
therefore Eqn.~(\ref{eq:i}) has to be further averaged to account for distribution of $\Omega$.
Using the normalization signal ($N_{\Omega} = \Delta\langle \frac{\Omega^3}{\Omega_A^3} c_1 s_1 s_2^2 \rangle_{L}  \exp \left( - 2 \tau/T_2\right)$), which is the SE signal with no pump pulse applied ($\langle \sin^2 \frac{\theta}{2} \rangle_L = 0$ in Eqn.~(\ref{eq:i})),
the analytical expression of the DEER spectrum ($I_{DEER} =\langle I_{\Omega} \rangle_{\Omega}/ \langle N_{\Omega} \rangle_{\Omega} $) is derived as,
\begin{equation}
\label{eq:deer}
\begin{split}
& I_{DEER} \left(\omega_B, [\omega_A, t_1, t_2, t_p, T, \{f\}_m, \{\omega\}_m], [\Omega, \Delta \omega, n] \right) \\
& = \frac {1}{\left\langle \left\langle S_A \right\rangle_{L} \right\rangle_{\Omega}}
\left\langle
\left\langle
S_A
\right\rangle_{L}
\exp\left(-\frac{2 \pi \mu_0 \mu_B^2 g_A g_B T n}{9\sqrt{3}\hbar} \left\langle S_B \right\rangle_{L}\right)
\right\rangle_{\Omega},
\end{split}
\end{equation}
where
\begin{eqnarray*}
S_A = \frac{\Omega^3}{\Omega_A^3} \cos(\Omega_A t_1/2) \sin(\Omega_A t_1/2) \sin^2(\Omega_A t_2/2)
\end{eqnarray*}
and
\begin{eqnarray*}
S_B = \frac{\Omega^2}{\Omega_B^2} \sin^2(\Omega_B t_p/2).
\end{eqnarray*}
$1/{\langle \langle S_A \rangle_{L} \rangle_{\Omega}}$ is the normalization factor.
$\left\langle...\right\rangle_{\Omega}$ denote averaging over the distribution of the Rabi frequency $\Omega$.
Among the arguments, in the DEER measurement, $\omega_B$ is variable, and $\omega_A$, $t_1$, $t_2$, $t_p$, $T$, $\{f\}_m$ and $\{\omega\}_m$ are fixed values.
Fitting parameters ($\Omega$, $\Delta \omega$ and $n$) are determined from analysis of the DEER spectrum as described in Sect.~IV-A.

\section{Discussion}
\subsection{Determination of N spin concentration}
In this section, we present the analysis of DEER spectrum to obtain the concentration of N spins.
The analysis was performed by fitting Eqn.~(\ref{eq:deer}) to the DEER signals.
In the case of the type-Ib diamond (Fig.~2b), the DEER pulse parameters ($t_1$ = 250 ns, $t_2$ = 450 ns, $t_p$ = 450  ns, $T$ = 2 $\mu$s, $\omega_A$ = 115 GHz) and the experimentally obtained $\{\omega\}_m$  (114.7714, 114.8008, 114.8865 and 114.9724 GHz) were used.
In addition, due to the magnetic field alignment along the $[111]$ crystallographic direction, the fraction of spins in each spectral line $f_m$ was set to $\{1/12, 3/12, 4/12, 3/12, 1/12\}$.
To account for the microwave field distribution, we used a sinusoidal function, $\Omega = \Omega_0 (1 + \cos (2 \pi x / \lambda_D))/2$, where $x$ is a distance of N spin from the surface of the diamond, $\lambda_D$ is the wavelength of the microwave in diamond ($\lambda_D$ = 1.08 mm at 115 GHz) and $\Omega_0$ is the maximum Rabi frequency in the diamond expressed in units of MHz, which was defined through the shortest duration of $\pi$ pulse ($t_{\Omega}$) in diamond as $\Omega_0 = 1/2 t_{\Omega}$.
Therefore, $\langle ... \rangle_{\Omega}$ in Eqn.~(\ref{eq:deer}) is equivalent to the averaging over the sample height $h$ (the dimension of the diamond sample along the magnetic field and $h$ = 2 mm in the present case).

\begin{figure}
\includegraphics[width=150 mm]{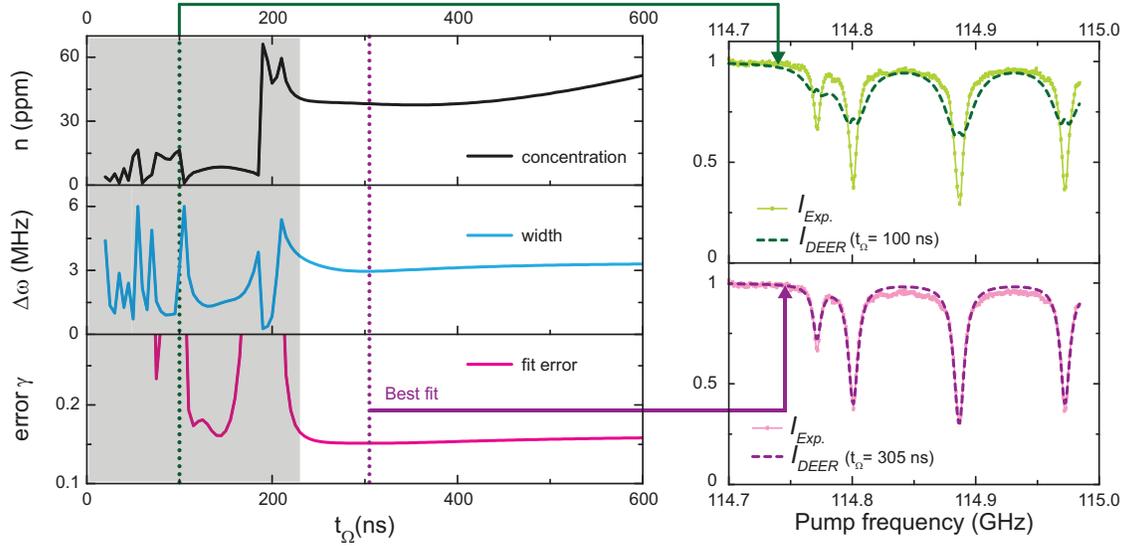}
\caption{Fit results of DEER spectrum.
Left top, middle, bottom panels show concentration $n$ of N spins, half width $\Delta \omega$ of inhomogeneous lineshape and a fit error $\gamma$, respectively, as obtained from the fit at a fixed $t_{\Omega}$ values.
The top (bottom) panel on the right shows the result of the fit obtained at $t_{\Omega}$ = 100 ns  (305 ns).
The result with $t_{\Omega}$ = 305 ns is the best fit.
The grey shaded area on the left indicates fits with a large $\gamma$.}
\label{fig:b}
\end{figure}
With the parameters defined above, we performed the fit of the experimental DEER spectrum $I_{Exp} (\omega_B)$ using a least squares minimization procedure with a fixed value of $t_{\Omega}$ and fitting parameters of $\Delta \omega$ and $n$.
The results of this procedure are shown in Fig.~\ref{fig:b} where $\Delta \omega$, $n$ and a fit error ($\gamma$) defined as a sum of squared residuals were plotted as a function of $t_{\Omega}$ ($t_{\Omega}$ = 20$-$600 ns).
We performed the fit in the wide range of $t_{\Omega}$ with a step size of 5 ns.
As seen in Fig.~\ref{fig:b}, the result of the fit highly depends on $t_{\Omega}$ and the fit error becomes smaller with $t_{\Omega}$ $\gtrsim$ 220 ns.
The minimum error value was obtained at $t_{\Omega}$ of 305 ns.
The values of $\Delta \omega$ and $n$ for the best fit (dashed violet line in Fig.~\ref{fig:b}) were obtained as 2.96$\pm$0.13 MHz and 38.2$\pm$0.8 ppm, respectively, where the error was calculated as 95 $\%$ confidence interval for the fit parameter.
Similarly, in the case of type-IIa diamond (Fig.~2b), the fit parameters were obtained as $t_{\Omega}$ = 300 ns, $\Delta\Omega$ = 0.49$\pm$0.14 MHz and $n$ = 0.14$\pm$0.01 ppm.
\begin{table}
\caption{Summary of $\Delta \omega$ and $n$ for the studied type-IIa and type-Ib diamonds as extracted from the analyses of the DEER data.}
\begin{center}
\begin{tabular} { c | c | c }
\hline\hline
$n$ (ppm) & $\Delta \omega$ (MHz) & $t_{\Omega} \pm$ 5 (ns) \\
\hline \hline
0.095$\pm$0.012 & 0.34$\pm$0.20 & 285 \\
0.139$\pm$0.011 & 0.49$\pm$0.14 & 300 \\
0.22$\pm$0.02 & 0.54$\pm$0.16 & 395 \\
0.26$\pm$0.03 & 0.40$\pm$0.16 & 460 \\
38.2$\pm$0.8 & 2.96$\pm$0.13 & 305 \\
22.4$\pm$0.4 & 2.36$\pm$0.12 & 110 \\
50.7$\pm$2.1 & 2.18$\pm$0.21 & 400 \\
86.1$\pm$0.8 & 3.93$\pm$0.26 & 370 \\
\hline
\end{tabular}
\end{center}
\end{table}
The fit results for all studied diamonds are summarized in Table~1.
The concentration for the shortest measured $T_2$ was found as 86.1$\pm$0.8 ppm, which is within the static model (Sec.~III-A).
The obtained $t_{\Omega}$ are consistent with the experiment where the lengths of the microwave pulses were chosen to maximize the SE signals (typical durations of the experimental $\pi$-pulses were on the order of a few hundreds of nanoseconds).
The values are also in a good agreement with our previous experiment~\cite{Stepanov15}.
Possible reasons for the variations are different sizes of the diamond crystals and imperfect sample positioning~\cite{Cho14}.

\subsection{$T_2$ vs N concentration}
\begin{figure}
\includegraphics[width=100 mm]{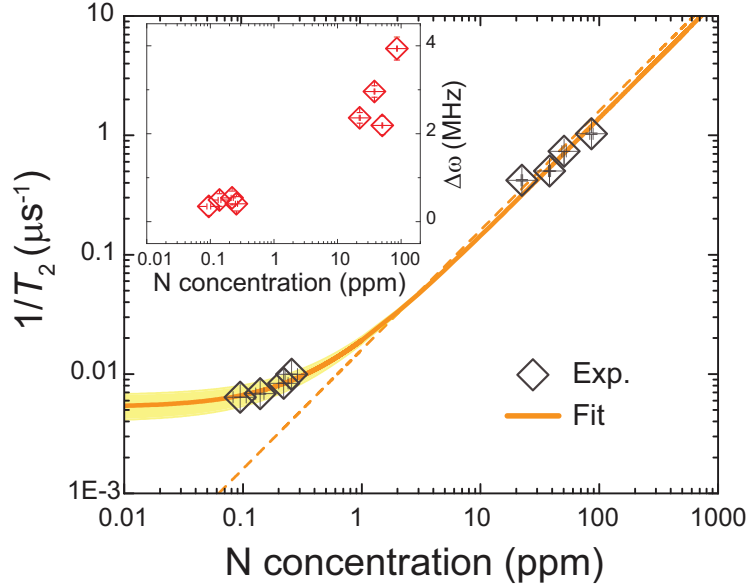}
\caption{
1/$T_2$ as a function of the N concentration.
Open squares represent experimentally obtained data, orange solid line is the best fit of the data to the model of decoherence rate described by Eqn.~\ref{eq:invT2}.
Yellow region represents the plot of Eqn.~\ref{eq:invT2} with the fixed $\Gamma_C$ in the range of 150$-$250 $\mu$s and a slope $C$=0.0139 $\mu$s$^{-1}$ppm$^{-1}$ as obtained from the best fit of the data.
Dashed orange line shows the best fit of the data using Eqn.~\ref{eq:invT2} without the nuclear spin decoherence ($1/T_{2}^{^{13}C}$ = 0).}
\end{figure}
Finally we discuss the relationship between $T_2$ and the concentration of N spins.
As shown in Fig.~6, $1/T_2$ increases while the N concentration increases in both type-Ib and type-IIa diamond.
In addition, the concentration dependence of the $1/T_2$ values are less pronounced in the type-IIa diamond.
To analyze the observed concentration dependence of $1/T_2$, we considered the two decoherence processes including the spin flip-flop process of N spins ($1/T_{2}^{N}$), where the contribution from the N spin is considered to be proportional to the N concentration ($1/T_{2}^{N} \sim n$), and the $^{13}$C decoherence ($1/T_{2}^{^{13}C}$).
Thus, the decoherence rate ($T_2$) is considered by,
\begin{eqnarray}
{\frac{1}{T_2}}=\frac{1}{T_{2}^{N}}
+\frac{1}{T_{2}^{^{13}C}} =C n
+\frac{1}{T_{2}^{^{13}C}},
\label{eq:invT2}
\end{eqnarray}
where $C$ is a proportional constant.
As shown in Fig.~6, the data is well explained with Eqn.~\ref{eq:invT2}.
From the fit using Eqn.~\ref{eq:invT2}, $C$ was estimated to be 0.0139$\pm$0.0005 $\mu$s$^{-1}$ppm$^{-1}$.
The N spin concentration dependence in $T_2$ was observed in type-Ib and natural type-Ia diamond crystals
although the previous study did not reveal the nuclear spin decoherence~\cite{Vanwyk97}.
Moreover, from the best fit, we estimated $T_{2}^{^{13}C}$ to be 190$\pm$10 $\mu$s.
This value is in a good agreement with the decoherence time due to $^{13}$C nuclear spins~\cite{Gaebel06, Takahashi08}.
In addition, we present the concentration dependence of the inhomogeneous linewidth ($\Delta \omega$).
As seen in the inset of Fig.~6, $\Delta \omega$ at the high concentrations (10$-$100 ppm) depends strongly on the concentration of N spins, suggesting that the linewidth is governed by the dipolar coupling between N spins.
In contrast, at the low concentrations ($<$ 1 ppm), the linewidth is almost independent of the concentration, suggesting that the broadening is dominated by other impurities, most probably $^{13}$C nuclear spins.

\section{summary}
In summary, we demonstrated the capability of 115 GHz DEER spectroscopy at room temperature to determine a wide range of N spin concentrations.
Using the pulsed 115 GHz ESR spectroscopy, we first determined $T_2$ in type-Ib and type-IIa diamond crystals and performed DEER spectroscopy to probe the magnetic dipole interaction between N spins.
From the analyses of the SE decay and the DEER spectra, we determined concentrations of N spins in the range of 0.1 $-$ 100 ppm with no reference sample.
Our DEER analysis to extract the spin concentration is strongly supported by the extracted N concentration dependence of the inhomogeneous linewidth and by the agreement of the estimated microwave power with our experimental values.
In addition, we showed that the measurement of the N spin concentrations allows us to determine contributions of N spins and $^{13}$C nuclear spins to $T_2$ quantitatively.
Moreover, the present methods is applicable to determine the concentration of NV ensembles and various other spin systems in solid.
In addition, by combining nanoscale magnetic resonance techniques based on NV centers,
this method may pave the way to determine spin concentrations within a microscopic volumes.

\section{acknowledgement}
This work was supported in part by the Searle Scholars Program, the USC Anton B. Burg Foundation and the National Science Foundation (DMR-1508661) (S.T.).


\end{document}